%% file: fpcp.tex
\documentclass[12pt]{article}
\usepackage{epsfig}

\input econfmacros.tex
\newcommand{\AmS}{{\protect\the\textfont2
  A\kern-.1667em\lower.5ex\hbox{M}\kern-.125emS}}
\newcommand{\pnn}{$K^+ \to \pi^+ \nu \bar\nu$}

\newcommand{\klpnn}{$K^0_L \to \pi^0 \nu \bar\nu$}
\newcommand{\klpnngen}{$K \to \pi \nu \bar\nu$}

\newcommand{\bklpnn}{$B(K^0_L \to \pi^0 \nu \bar\nu)$}
\newcommand{\brklpnn}{$B(K^0_L \to \pi^0 \nu \bar\nu)$}
\newcommand{\bkpnn}{$B(K^+ \to \pi^+ \nu \bar\nu)$}
\newcommand{\kpnngen}{$K \to \pi \nu \bar\nu$}

\def\epp{$\epsilon^{\prime}$}
\def\vep{$\epsilon$}

\def\be{\begin{equation}}
\def\ee{\end{equation}}
\def\bea{\begin{eqnarray}}
\def\eea{\end{eqnarray}}

\def\vtd{V_{td}}

\def\Title#1{\begin{center} {\Large {\bf #1} } \end{center}}

\begin{document}

\Title{Rare Kaon Decays: Progress and Prospects}

\bigskip\bigskip


\begin{raggedright}  

{\it Douglas Bryman\index{}\\
Department of Physics and Astronomy\\
        University of British Columbia \\
        6224 Agricultural Road, Vancouver, B.C. V6T 1Z1, Canada}
\bigskip\bigskip
\end{raggedright}

\section{Introduction}

There is a large overlap in the current physics motivation for
investigating B-meson and K-meson decays. The much newer and heavier b
quark opens many possibilities for studying interesting decay channels
with fascinating phenomenologies  of mixing,
oscillations, and CP violation. In the past year the promise of
revealing the origin of CP violation has begun to be realized in
pioneering experimental studies at $e^+ e^-$ colliders.  The
asymmetries observed in $B\rightarrow \Psi K_S$ decays by BABAR and
BELLE provide the first conclusive evidence for CP violation outside
the neutral K system.
  
K decays, a much older game, have long been a gold mine of surprising
information at the forefront of particle physics. K decay experiments opened
the doors to quark mixing and  CP violation. In the
intervening years, the effort to establish consistency with the
Standard Model (SM) description involving direct CP violation was a
protracted battle. Nevertheless, KTEV\cite{kteve} and NA48\cite{na48e}
finally agreed on evidence for direct CP violation through consistent
non-zero values of $\epsilon'/\epsilon$ in $K\rightarrow \pi \pi$
decays. However, as with many approaches to CP violation involving K
and B decays, the underlying short distance physics is not easy to get
at due to complications associated with strong interactions
including penguin diagram processes.

The CP-violating asymmetry in $B\rightarrow \Psi K_S$ decays which
determines $sin(2\beta)$, and the ratio of $B_s/B_d$ mixing
($x_s/x_d$) which yields $|\frac{V_{ts}}{V_{td}}|$ stand out as
theoretically unambiguous  quantities, relatively free from strong
interaction uncertainties. Together they can cleanly determine the apex of the
usual unitary triangle giving a powerful test of the consistency of
the SM. $sin(2\beta)$ is becoming known to increasing precision as
discussed at this conference and there is optimism that $B_s $ mixing
will be measured in Run II at the Tevatron, although extracting
$|V_{td}|$ is subject to some uncertainty due to SU(3)-breaking
effects.
   
During the past few years the state-of-the-art in rare
K decays has reached  single event sensitivities of $10^{-12}$ in
experiments at the Brookhaven National Laboratory (BNL) Alternating
Gradient Synchrotron (AGS). BNL E871 reported 6200 events of
$K^0_L\rightarrow \mu^+\mu^-$ decay at the $10^{-8}$ level\cite{kmm}
and 4 events of $K^0_L\rightarrow e^+e^-$ at $10^{-11}$\cite{kee}. Two
events were observed by BNL E787 with the kinematically incomplete \pnn\
signature at the $10^{-10}$ level\cite{pnn02}. These successes have
opened up the possibility of fully accessing the ``dynamic duo'' of
\pnn\ and \klpnn\ which offer the best possibilities for obtaining
high quality information on short distance physics, complementary and
comparable in precision to the leading approaches using B mesons.

The observation of \pnn\ by E787 is consistent with the SM 
expectation. To fully explore the possibility of new physics or to
make a precise measurement of the t-d quark coupling $|\vtd |$,
a new measurement has just
commenced. E949 at the BNL AGS is designed to obtain sensitivity an
order of magnitude below the SM prediction. Later in the decade, the
CKM experiment\cite{ckm} at Fermilab will begin to pursue \pnn\ at
even higher precision.

\begin{figure}[htb]
\begin{center}
\epsfig{file=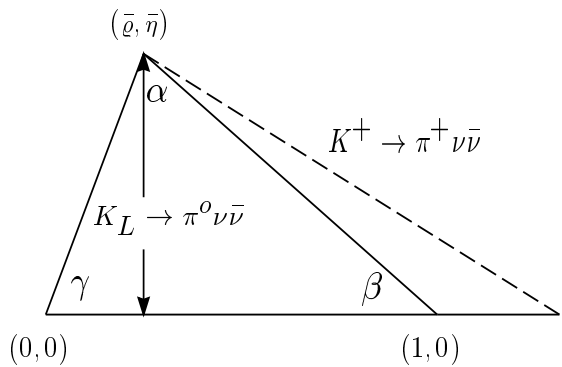,width=3in,height=2in,clip=t}
\caption{\label{triangle} The unitarity triangle.}
\end{center}
\end{figure}

The $K$ sector can also yield the
single most incisive measurement in the study of CP violation through
a measurement of the branching ratio for \klpnn\ decay.  Within the SM
context, this is a unique quantity which directly measures the area of
the CKM unitarity triangles {\it{i.e.}} the physical parameter that
characterizes all CP violation phenomena, or the height of the
triangle shown in Fig.~\ref{triangle}.  The quest to observe \klpnn\
is beginning in earnest with a pilot experiment at KEK (E391A) and the
new KOPIO experiment at BNL.
The measurements of \bkpnn\ and \bklpnn\ will result in a complete
picture of SM CP violation in the $K$ system and a comparison with
comparably precise measurements anticipated from the $B$ sector will
be possible.

Other areas of current activity in the field of rare K decays include studies
of radiative processes relevant to Chiral Perturbation Theory (CHPT)
and searches for exotic reactions. 

\section{THEORY OF $K \to \pi \nu \overline\nu$ }

\begin{figure}
\begin{center}
\epsfig{file=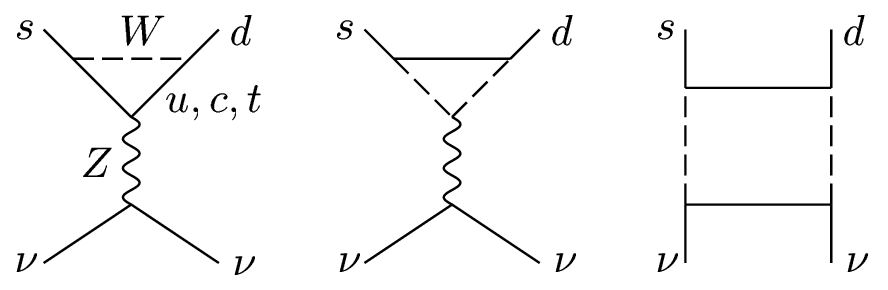,width=3in,height=0.86in, clip=t}
\caption{\label{feyndiag} The leading electroweak diagrams
inducing \klpnngen~ decays. For \klpnn~ only the top quark contributes.}
\end{center}
\end{figure}

While clean extraction of SM parameters from $K$ decay observables
like \epp/\vep\ is presently precluded due to non-perturbative strong
interaction uncertainties, \pnn\ and ~\klpnn~ do not suffer from these
maladies.  In \pnn\ and ~\klpnn\ decays, which arise at the one loop
level in the SM as shown in Fig.~\ref{feyndiag}, hadronic effects are
well known from measurements of the similar decays K$\to\pi e \nu$
related by isospin\cite{Marciano96}. The presence of the top quark in
the loops makes these decays very sensitive to $\vtd$~\cite{Buras97}
and the  simple final states allow unusually precise calculations to be
made. Small QCD corrections have been calculated to
next-to-leading-logarithmic order~\cite{Buchalla94}. Long distance
effects are negligible in comparison with short distance effects.
\pnn\ is expected to occur with a branching ratio of $0.72\pm 0.21
\times 10^{-10}$\cite{isidori} with a purely theoretical uncertainty
of about 7\% due to a charm quark contribution to the loops.

\klpnn~ decay is unique in that it is completely dominated by
direct CP violation~\cite{Littenberg89}. Since $K^0_L$ is
predominantly a coherent, CP odd superposition of $K^0$ and $\bar
K^0$, only the imaginary part of $V^*_{ts} V_{td}$ survives in the
amplitude.  The lack of a significant charm quark
contribution reduces the intrinsic theoretical uncertainty to
$\cal{O}$$(2\%)$.  Since the value of the sine of the Cabibbo angle is
well known, $Im(V^*_{ts} V_{td})$ is
equivalent to the Jarlskog invariant, $\cal{J}$$ \equiv - Im(V_{ts}^*
V_{td} V_{us}^* V_{ud}) = -\lambda (1 - \frac{\lambda^2}{2})
Im(V_{ts}^* V_{td})$.  $\cal{J}$ is  equal to twice the area
of any of the six possible unitarity triangles~\cite{Jarlskog}.
Since theoretical uncertainties are extremely small, measurement of
\bklpnn\ will provide the standard against which all other measures of
CP violation will be compared, and even small deviations from the
expectation derived from SM predictions or from other measurements,
{\it e.g.} in the $B$ sector, will unambiguously signal the presence
of new physics.  In the SM context the branching ratio for \klpnn\ is
given by\cite{Buras99}
\begin{equation}\label{bklpnw}
B(K^0_L\to\pi^0\nu\bar\nu)=1.8 \cdot 10^{-10}(\frac{Im(V_{ts}^* V_{td})}{\lambda^5})^2  X^2(x_t)=1.8 \cdot 10^{-10} \eta^2 A^4 X^2(x_t),
\end{equation}
where $X(x_t)$ is a kinematic function of the top quark mass.  The
branching ratio for \klpnn~ is expected to lie in the range $(2.6 \pm
1.2)\cdot 10^{-11}$.  A clean measure of the height of the unitary
triangle, $\eta$, is provided by the \klpnn~ branching ratio.  All
other parameters being known, Eq. \ref{bklpnw} implies that the
relative error of $\eta$ is half that of a measurement of \bklpnn.
Thus, for example, a $15\%$ measurement of \bklpnn~ can, in principle,
determine $\eta$ to $7.5\%$.

Most forms of new physics~\cite{Leurer93,Grossman97,isidori}
postulated to augment or supersede the SM have implications for
B(\kpnngen).  In minimal supersymmetry and in some multi-Higgs doublet
models~\cite{Belanger92}, the extraction of $\sin 2\alpha$ and $\sin
2\beta$ from CP asymmetries in B decays would be unaffected.  Such
effects might then show up in a comparison with \klpnn, where, e.g.,
charged Higgs contributions modify the top quark dependent function in
\bklpnn.  In other new physics scenarios, such as supersymmetric
flavor models~\cite{Nir98}, the effects in \kpnngen~ tend to be small,
while there can be large effects in the $B$ (and also the $D$) system.
In these models the rare $K$ decays are the only clean way to measure
the true CKM parameters.  Examples of new physics scenarios that show
large deviations from the SM in \kpnngen\ are provided by some
extended Higgs models, in topcolor-assisted technicolor models
\cite{Xiao99}, in left-right symmetric models~\cite{Kiyo99}, in models
with extra quarks in vector-like representations~\cite{Grossman97},
lepto-quark exchange \cite{Grossman97}, and in 4-generation models
\cite{Hattori99}.  The confirmation of a relatively large value for
$\epsilon'/\epsilon$ has focussed attention on the
contributions of flavor-changing $Z$-penguin diagrams in generic
low-energy supersymmetric extensions of the SM~\cite{Buras99}.  Such
diagrams can interfere with the weak penguins of the SM, and either
raise or reduce the predicted \brklpnn~ by considerable factors. The
effects of SUSY and other non-SM approaches on the $K$ and $B$ system
generally turn out to be  different and apparent
discrepancies between measurements of SM quantities would be indications
of new physics.

\section{Measurement of \pnn}

The final results from BNL E787 reported recently\cite{pnn02} were
based on observations of decays of $5.9 \times 10^{12}$ K mesons
at rest.  Measurement of the $K^+ \! \rightarrow \! \pi^+ \nu
\overline{\nu}$ decay involves only the $\pi^+$ track and $\pi^+$
decay products and was accomplished with an efficiency of $2 \times
10^{-3}$. The efficiency is relatively small because of the necessity to
suppress similar background processes by up to 10 orders of magnitude.
Major background sources include the two-body decays $K^+
\!  \rightarrow \! \mu^+ \nu_\mu$ and $K^+ \!
\rightarrow \!  \pi^+ \pi^0$, pions scattered from the
beam, and $K^+$ charge exchange (CEX) reactions resulting in decays
$K_L^0\to\pi^+ l^- \overline{\nu}_l$, where $l=e$ or $\mu$. In order
to make an unambiguous measurement of $B(K^+ \to \pi^+ \nu \bar\nu)$,
the sum of all backgrounds was suppressed to the estimated level of
$0.15^{+0.05}_{-0.04}$ events. The background suppression procedures and
estimates were subjected to extensive verification, and a
signal-to-background function was created to evaluate potential
\pnn\ candidates.

\begin{figure}[htb]
\begin{center}
\epsfig{file=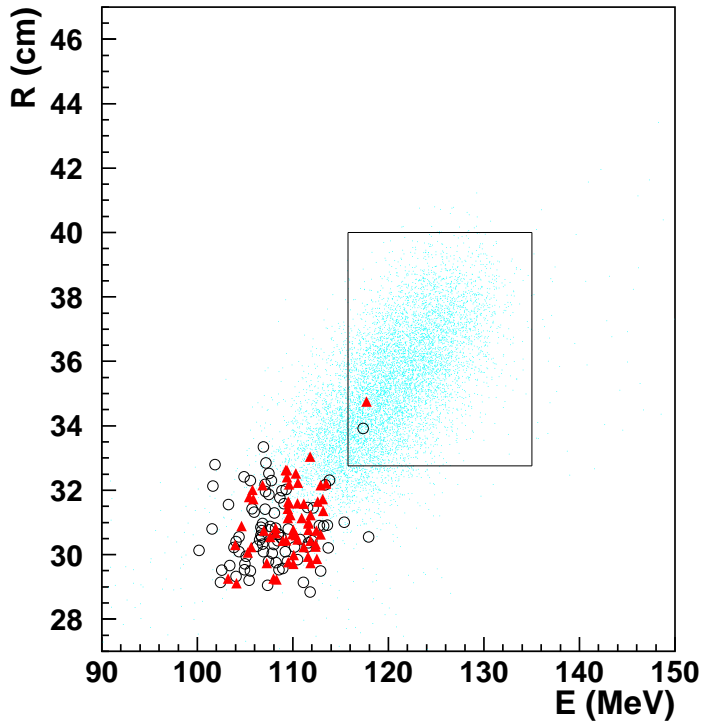,height=4in}
\caption{ Range  vs. energy plot of the E787
final sample for \pnn. The circles are for the 1998 data and the triangles are
for the 1995-97 data set. The group of events around $E=108$~MeV is
due to the $K^+\rightarrow \pi^+ \pi^0$ background.  The simulated distribution of
expected events from $K^+ \!  \rightarrow \!  \pi^+ \nu
\overline{\nu}$ is indicated by dots.}
\label{rve}
\end{center}
\end{figure}

The combined result for E787 data taken between 1995 and 1998 is
shown in Fig.~\ref{rve}, the range vs. kinetic energy of events
surviving all other cuts.
In Fig.~\ref{rve} the box represents the signal region in which two
events appear.  A likelihood ratio technique\cite{Junk} was used to
determine the best estimate of the branching ratio.  Based on two
observed events and the expected background levels, the result is
$B(K^+ \!  \rightarrow \!  \pi^+ \nu \overline{\nu}) =
1.57^{+1.75}_{-0.82} \times 10^{-10}$.  This result is consistent with
the SM prediction. The estimated probability for the observation  to be
due entirely to background is at the level of
0.02\% \cite{Junk}. D'Ambrosio and Isidori\cite{isidori} have illustrated
the impact of the current E787 results on the measurements of SM quark
mixing and CP violation parameters and discussed the possibilities for
identifying new physics.

The follow-on experiment E949 is an improved  version of E787.
Upgrades included enhancements to the photon detection
system, and improvements to the DAQ system, beam counters, tracking
chamber electronics, monitoring systems, among others. E949 is aiming
for 
a single event sensitivity of approximately $10^{-11}$,
an order of magnitude below the SM level. It  began
data-taking in early 2002.

\begin{figure}[htb]
\begin{center}
\epsfig{file=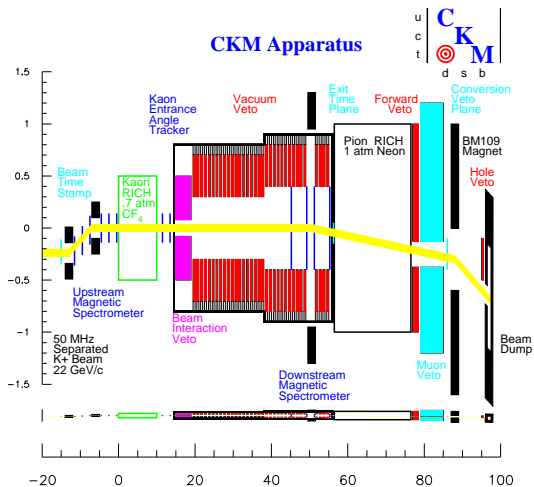,height=3in}
\caption{Layout of the proposed CKM experiment at Fermilab.}
\label{ckm}
\end{center}
\end{figure}

In the longer term future, the proposed CKM experiment\cite{ckm} at
Fermilab shown schematically in fig.~\ref{ckm}
seeks another order of magnitude in sensitivity, i.e.
$10^{-12}$. CKM uses particle identification based on ring imaging
Cerenkov detectors along with momentum measurements to suppress
backgrounds. At the SM level, a signal of approximately 95 events with
a signal/noise ratio $S/N= 7$ is sought.

\section{Prospects for Measuring \klpnn}

The experimental challenges of studying B$(K^0_L \rightarrow \pi^0
\nu\bar{\nu})$ are comparable to those encountered in the measurement
of \pnn.   There are similar competing decays, particularly
$K_L^0\rightarrow \pi^0 \pi^0$, that also yield $\pi^0$s but with branching
ratios that are millions of times larger.
Interactions between neutrons and kaons in
the neutral beam with residual gas in the decay volume can also result
in emission of single $\pi^0$s, as can the decays of hyperons which
might occur in the decay region. The
current experimental limit \bklpnn$< 5.9 \times
10^{-7}$~\cite{Alavi-Harati00} was a by-product of the KTeV experiment
at Fermilab.

To definitively measure \klpnn, a detection technique must be
developed that provides maximum possible redundancy for this
kinematically unconstrained decay, that has an optimum system for
insuring that the observed $\pi^0$ is the only detectable particle
emanating from the $K^0_L$ decay, and that has multiple handles for
identifying possible small backgrounds that might simulate the desired
mode.

A first step at a dedicated search for \klpnn~ is commencing soon at
KEK with experiment E391a which aims for a sensitivity of
$10^{-10}-10^{-9}$. This experiment uses a narrowly collimated
``pencil'' neutral beam\cite{Watanabe} along with a high acceptance
hermetic detector. E391a 
will test the limits of reliance on photon detection efficiency to
suppress the major backgrounds and may lead to a proposal at the
emerging Japanese Hadron Facility.

The new KOPIO experiment at BNL is scoped to reach a sensitivity more
than an order of magnitude below the SM prediction, i.e. $10^{-12}$,
aiming for at least 50 events of \klpnn~ with $ S/N > 2$.  KOPIO, which
is presently in the R\&D and design phase, employs a low momentum
time-structured $K^0_L$ beam to allow determination of the incident
kaon momentum by time-of-flight.  This intense beam, with its special
characteristics, can be provided by the BNL AGS.  Utilizing low
momentum also permits a detection system for the $\pi^0$ decay photons
that yields a fully constrained reconstruction of the $\pi^0$'s mass,
energy, and, momentum.  The system for vetoing extra particles is well
understood. These features which are similar to those employed
successfully in the E787 measurement of \pnn\ provide the necessary
redundancy and checks.
\begin{figure*}[htpb]
    \begin{minipage}{0.38\linewidth}
    \centerline{\epsfig{figure=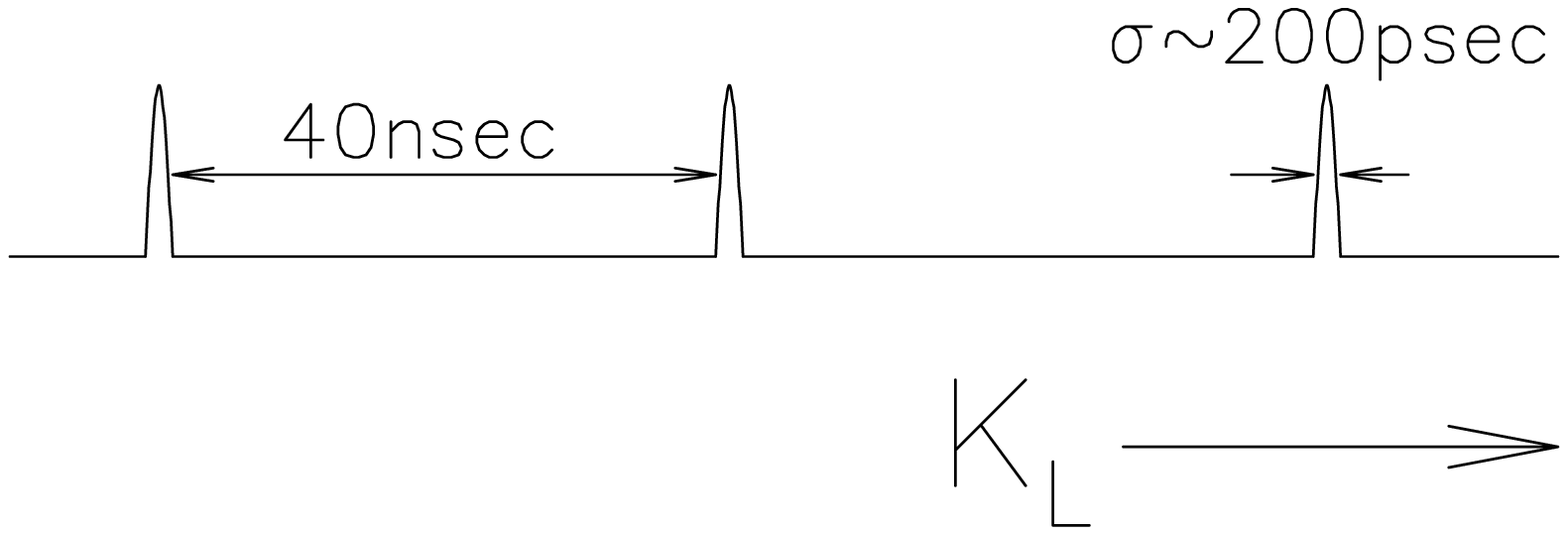,height=1.5in,width=2in}}
    \vspace{1in}
    \end{minipage}\hfill
\begin{minipage}{0.52\linewidth}
    \centerline{\epsfig{figure=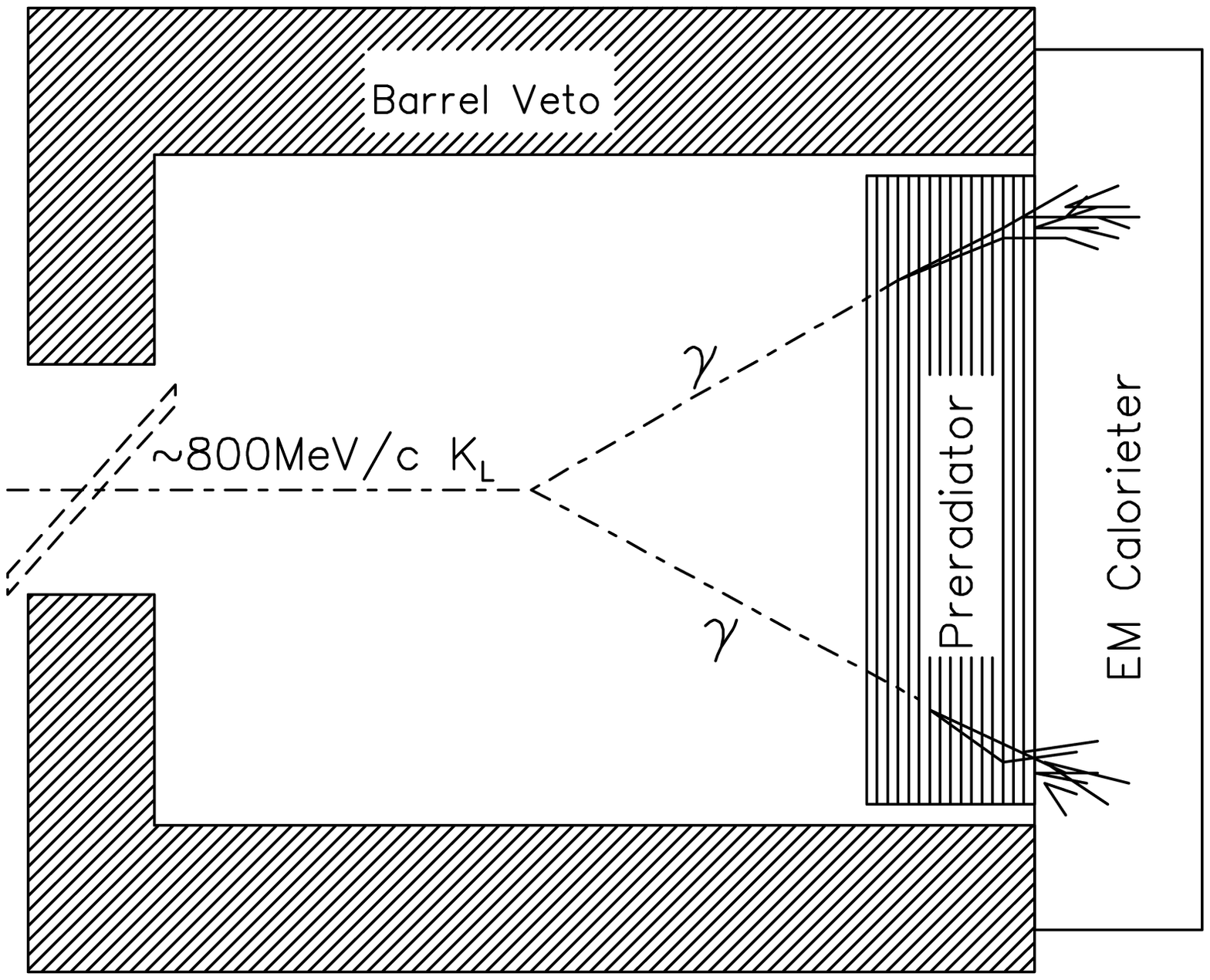,height=2in,width=3in}}
    \end{minipage}\hfill
    \caption{Elements of the KOPIO concept : a  pulsed primary beam
produces low energy kaons whose time-of-flight reveals their
momentum when the $\pi^0$ from \klpnn~ decay is reconstructed. }
    \label{f:schem}
\end{figure*}

Figure~\ref{f:schem} shows a simplified representation of the beam and
detector concept for KOPIO.  Photons from \klpnn~ decay are observed in a
two-stage endcap detector comprised of a fine-grained preradiator with
angular resolution approximately 25 mr for photons of 250 MeV. It is
followed by a shashlyk-type electromagnetic calorimeter.  The
preradiator-calorimeter combination is expected to have an energy
resolution of $\sigma_{\rm E}$/E$< 0.03/\sqrt{{\rm E(GeV)}}$. The
entire decay region is surrounded by efficient charged particle and
photon detectors.

\section{Other Rare K Decays}

Considering the obvious difficulty of measuring \klpnn~ there is still
interest in studying $K_L^0\rightarrow \pi^0 e^+ e^-$ to obtain
information on direct CP violation. However this decay presents
formidable experimental and theoretical challenges.  The direct
CP-violating component which is due to diagrams similar to those for
\klpnn~ is estimated to occur at a branching ratio of $4\times
10^{-12}$. There are potentially important competing effects from
CP-conserving amplitudes, CP-violating amplitudes due to mixing, and
backgrounds from similar decay signatures like $K_L^0\rightarrow
\gamma \gamma e^+ e^-$\cite{LL-PDG}.

The present limit is B($K_L^0\rightarrow \pi^0 e^+ e^-$)$<5.1 \times
10^{-10}$\cite{klpee}. The NA48 collaboration has recently reported a
result on the branching ratio and shape of the spectrum for
$K_L^0\rightarrow \pi^0 \gamma\gamma$ which can be used to limit the
CP-conserving component of $K_L^0\rightarrow \pi^0 e^+ e^-$\cite{NA48}. A total
of 2558 $K_L^0\rightarrow \pi^0 \gamma\gamma$ events were observed
with a background estimated to be 3.2\% giving the results
B($K_L^0\rightarrow \pi^0 \gamma\gamma$)$=
1.36 \pm 0.03_{stat} \pm 0.03_{syst} \pm 0.03_{norm} \times 10^{-6}$ and the
vector coupling constant $a_v= -0.46 \pm 0.03_{stat} \pm 0.04_{syst}$.
\footnote{Comparable measurements reported by the KTEV
collaboration\cite{KTEVklpgg} gave B($K_L^0\rightarrow \pi^0
\gamma\gamma$)$= 1.68\pm 0.10 \times 10^{-6}$ and $a_v= -0.72 \pm 0.05
\pm 0.06$.} Using these results and a limit on the rate in the region
$m_{\gamma\gamma}<m_{\pi^0}$, NA48 presented an estimate for the
CP-conserving component of B($K_L^0\rightarrow \pi^0 e^+
e^-$)$_{CPC}=4.7^{+2.2}_{-1.8}\times 10^{-13}$, suggesting that it
may  not dominate in the measurement of CP-violating effects in
$K_L^0\rightarrow \pi^0 e^+ e^- $. However, there is also a
dispersive amplitude which is not reliably calculated which could be
of comparable importance\cite{LL-PDG}.

The NA48 experiment has now been upgraded to
NA48/1\cite{sacco_moriond}. Improvements made to the
detectors and beam lines will enable experiments with 100x (or greater)
intensities to achieve single event sensitivities to rare decays at
the $10^{-10}$ level. NA48/1 plans to study many processes including
$K_S^0\rightarrow \pi^0 l^+ l^-$ (which can be used to limit the
contribution to indirect CP violation in $K_L^0\rightarrow \pi^0 e^+ e^-
$), $K_S^0\rightarrow 3\pi$, and time-dependent CP-violating effects
in $K_{L,S}^0\rightarrow \pi \pi \gamma^*$ decays.  In addition,
semi-leptonic and radiative neutral hyperon decays will be
investigated.

\section{Summary and Prospects}

Rare kaon decay experiments underway or planned for the BNL AGS,
Fermilab and KEK aim to study the extraordinary decays \pnn\ and
\klpnn.  BNL E787 has presented evidence for \pnn\ at the $10^{-10}$
level based on the observation of two clean events leading the way for
BNL E949 and CKM at Fermilab which seek one and two orders of
magnitude greater sensitivities. The proposed KOPIO measurement of
\bklpnn\ will allow a determination of the imaginary part of
$V^*_{ts}V_{td}$ giving the fundamental CP-violating parameter of the
SM, in a uniquely clean manner.  Since the measurement of
\bkpnn\ determines $|V^*_{ts} V_{td}|$, a complete derivation of the
unitarity triangle is facilitated. The parameters derived from the $K$
decays  will be compared to high precision data expected to come from
the $B$ sector allowing for incisive searches for new physics.

New results on several other rare K decays have recently become
available from NA48 ($K_L^0\rightarrow \pi^0 \gamma\gamma$), the
HYPER-CP collaboration ($K^{\pm}\rightarrow \pi^{\pm} \mu \mu$)\cite{hypercp}
and KLOE at DAPHNE ($K_S^0\rightarrow \pi e \nu$)\cite{KLOE} shedding
light on issues in CP violation and Chiral Perturbation Theory
(CHPT). Other new results on exotic processes have been presented
recently by BNL E787 ($K^+\rightarrow \pi^+ a$\cite{kpx} and
$K^+\rightarrow \pi^+ \gamma$\cite{kpg}) and new results on lepton
flavor violating decays and other processes
are  anticipated from E865 at BNL and KTEV/E799 at Fermilab.

At CERN and DAPHNE, additional rare radiative decays of $K_L$ and $K_S$ are being
vigorously pursued along with many other studies relevant to tests of
CP/CPT violation and CHPT by the upgraded NA48/1 experiment and by
KLOE.

\section{Acknowledgements} I would like to thank L. Littenberg for
comments on the manuscript.

\end{document}

%% file: econfmacros.tex
\def\beq{\begin{equation}}
\def\eeq#1{\label{#1}\end{equation}}
\def\eeqn{\end{equation}}

\def\beqa{\begin{eqnarray}}
\def\eeqa#1{\label{#1}\end{eqnarray}}
\def\eeqan{\end{eqnarray}}

\let\bar=\overbar

\def\Dslash{\not{\hbox{\kern-4pt $D$}}}
\def\dslash{\not{\hbox{\kern-2pt $\del$}}}

\def\ee{e^+e^-}

\def\msb{{\bar{\ssstyle M \kern -1pt S}}}

